\documentclass[12pt]{article}
\usepackage{a4wide,amssymb,cite,slashed}
\pdfoutput=1

\usepackage{a4wide,amssymb,cite,graphicx}
\usepackage{epsfig}
\usepackage[usenames,dvipsnames]{color}
\usepackage{slashed}
\newcommand{\be}{\begin{equation}}
\newcommand{\ee}{\end{equation}}
\newcommand{\bea}{\begin{eqnarray}}
\newcommand{\eea}{\end{eqnarray}}

\def\circa#1{\,\raise.3ex\hbox{$#1$\kern-.75em\lower1ex\hbox{$\sim$}}\,}

\begin{document}

\begin{titlepage}
%
%


%

\begin{centering}
\vspace{1cm}
{\Large {\bf A minimal flavored $U(1)'$ for $B$-meson anomalies}} \\

\vspace{1.5cm}

{\bf  Ligong Bian$^{*,\dagger}$, Soo-Min Choi$^\dagger$, Yoo-Jin Kang$^{\dagger,\ddagger}$ and Hyun Min Lee$^\dagger$}
\vspace{.5cm}

{\it $^*$Department of Physics,
Chongqing University, Chongqing 401331, China. }
\\
{\it $^\dagger$Department of Physics, Chung-Ang University, Seoul 06974, Korea.} 
\\
{\it $^\ddagger$Center for Theoretical Physics of the Universe, Institute for Basic Science, 
   \\ Daejeon, 34051, Korea. }

\end{centering}
\vspace{2cm}

\begin{abstract}
\noindent
We consider an anomaly-free $U(1)'$ model with favorable couplings to heavy flavors in the  Standard Model(SM), as motivated by $B$-meson anomalies at LHCb. Taking the $U(1)'$ charge to be $Q'=y(L_\mu-L_\tau)+ x(B_3-L_3)$, we can explain the $B$-meson anomalies without invoking extra charged fermions or flavor violation beyond the SM.
We show that there is a viable parameter space with a small $x$ that is compatible with other meson decays, tau lepton and neutrino experiments as well as the LHC dimuon searches.  We briefly discuss the prospects of discovering the $Z'$ gauge boson at the LHC in the proposed model.

\end{abstract}

\vspace{3cm}

\end{titlepage}

\section{Introduction}


Anomalies in rare semi-leptonic decays of $B$-meson have been recently reported in the observables such as $R_K$ \cite{RK}, $R_{K^*}$ \cite{RKs}, $P'_5$ \cite{P5}, etc. 
In particular, there is a $2.6\sigma$ discrepancy from the SM in $R_K={\cal B}(B\rightarrow K\mu^+\mu^-)/{\cal B}(B\rightarrow Ke^+e^-)$ \cite{RK}, as follows,
\bea
R_K=0.745^{+0.097}_{-0.082}, \quad 1\,{\rm GeV}^2<q^2 <6\,{\rm GeV}^2, 
\eea
and there are $2.1-2.3\sigma$ and $2.4-2.5\sigma$ discrepancies in the decay mode with vector meson, $R_{K^*}={\cal B}(B\rightarrow K^*\mu^+\mu^-)/{\cal B}(B\rightarrow K^*e^+e^-)$ \cite{RKs}, as follows,
\bea
R_{K^*}&=& 0.66^{+0.11}_{-0.07}(\rm stat)\pm 0.03({\rm syst}), \quad 0.045\,{\rm GeV}^2<q^2 <1.1\,{\rm GeV}^2,  \nonumber \\
R_{K^*}&=& 0.69^{+0.11}_{-0.07}({\rm stat})\pm 0.05({\rm syst}), \quad 1.1\,{\rm GeV}^2<q^2 <6.0\,{\rm GeV}^2.
\eea
As the SM predictions for $R_K$ or $R_{K^*}$ are almost unity, if confirmed, the above anomalies would be a hint at new physics for lepton non-universal interactions \cite{muon,electron,fits,models}, such as flavor-dependent $Z'$ \cite{C9,schmaltz,yavin1,Zprime,flavor,ko,bhatia,yanagida,tang} and its non-abelian extensions \cite{su2}, leptoquarks \cite{schmaltz,rpv,leptoquark}, light mediators \cite{lightZprime}, extra charged fermions \cite{heavyfermion}, etc. On the other hand, we may need more caution in interpreting the anomalies in the angular distributions of $B$-meson decays such as $P'_5$, because it is prone to hadronic power corrections \cite{qcd}.  

In this article, we consider a minimal extension of the Standard Model (SM) with extra $U(1)'$ gauge symmetry as a linear combination of $U(1)_{L_\mu-L_\tau}$ and $U(1)_{B_3-L_3}$. 
In this model, we need only two right-handed neutrinos to cancel the gauge anomalies but it is also necessary  to  include one more right-handed neutrino with zero $U(1)'$ charge for neutrino masses and mixings.  Furthermore, we extend the Higgs sector by including extra Higgs doublet and three complex singlet scalars and obtain the predictive patterns of quark and lepton mass matrices that are consistent with experimental data. There are similar minimal $U(1)'$ models for explaining the $B$-meson anomalies and neutrino mass generation where there are three right-handed neutrinos, extra Higgs doublet and scalar singlet(s) beyond the SM \cite{flavor,bhatia}. 

In the proposed model, $U(1)_{L_\mu-L_\tau}$ violates lepton universality such that $Z'$ decays favorably into muons. Moreover, $U(1)_{B_3-L_3}$ gives rise to $b$ to $s$ flavor changing interaction for $Z'$ in combination with the CKM mixing in the SM, but without a need of new flavor violation. 
Therefore, we can accommodate the observed ratios of $R_K$ and $R_{K^*}$ easily, avoiding other experimental constraints.   
Imposing the bounds from the $B_s-{\bar B}_s$ mixing and the dimuon $Z'$ searches at the LHC, we  need to choose the mixing with $U(1)_{B_3-L_3}$ in $U(1)'$ to be small enough. Then, we show that there is a viable parameter space for explaining the $B$-meson anomalies, being consistent with bounds from other meson decays, tau lepton, neutrino experiments and electroweak precision data as well as LHC dimuon searches.  We also discuss how the model is constrained by existing experiments and it will be tested with upcoming data from the LHC.

Given that there are similar $U(1)'$ models in the literature, we compare them to our model and point out the main differences.  In particular, the mixings between  $L_\mu -L_\tau$ and other anomaly-free symmetries have been discussed. For instance, a mixing between  $L_\mu -L_\tau$ and  $B_1+B_2-2B_3$ \cite{flavor} or $B_3-L_1$ \cite{ko} was proposed to explain the $B$-meson anomalies in a similar way, but the resulting flavor structure in the lepton sector can be different. A similar model based on $U(1)_{B_3-L_3}$ as ours has been proposed \cite{yanagida} but the $Z'$ coupling to muon is induced by the flavor mixing of the lepton sector in this case \footnote{See also Ref.~\cite{BL3} for the phenomenology of a light $Z'$ of mass below multi-GeV scale.}. Furthermore, there are general discussions on anomaly-free combinations of flavor-dependent $U(1)'$ symmetries with lepton and baryon numbers without introducing extra chiral fermions \cite{tang}, but there is no phenomenological discussion on our model up to now.

The paper is organized as follows. 
We first introduce the model with extra $U(1)'$ and construct the Lagrangian of the model. 
Then we constrain the parameter space to explain the $B$-meson anomalies in our model and discuss the bounds from $B_s-{\bar B}_s$ mixing, other meson decays, $(g-2)_\mu$, tau decays, neutrino trident production, electroweak precision data as well as the LHC dimuon searches. 
Finally, conclusions are drawn.

\section{The Model}

Interesting candidates for anomaly-free $U(1)'$ symmetries  beyond the Standard Model  are $U(1)_{L_e-L_\mu}$, $U(1)_{L_\mu-L_\tau}$ and $U(1)_{L_\tau-L_e}$, with no extra fermions, and $U(1)_{B-L}$ with three right-handed neutrinos. The $U(1)_{L_\mu-L_\tau}$ symmetry violates lepton universality as required by $B$-meson anomalies at the LHCb, but it couples only to leptons. On the other hand, $U(1)_{B-L}$ is anomaly-free in each generation of quarks and leptons and well motivated by gauge coupling unification such as $SO(10)$ GUT. But, $U(1)_{B-L}$ couples to both electron and muon equally and it is highly constrained by dijet or dilepton bounds at the LHC. So, the new $b$ to $s$ transition induced by the SM-like loop processes with $U(1)_{B-L}$  is too small to explain the $B$-meson anomalies for $g_{B-L}\sim g_Y$. 
Instead,  we can define $U(1)_{B_i-L_i}$ with $i=1,2,3$ per generation, each of which is anomaly-free with one right-handed neutrino.    

The generation-dependent $U(1)_{B-L}$ could be realized in a UV theory with enhanced gauge symmetries that are broken down into a generation-dependent $U(1)_{B_i-L_i}$ at low energy. For instance, in a gauged $U(1)$ clockwork theory for $U(1)_{B-L}$ where there are multiple copies of independent $U(1)_{B-L}$ gauge bosons, the localization of zero mode of the gauged $U(1)_{B-L}$ could explain the generation-dependent $B-L$ charges as well as the fractional $B-L$ charges depending on the localization of third generation fermions \cite{u1cw}. In this case, the extra massive $U(1)_{B-L}$ gauge bosons have a mass gap such that there is no sizable impact on the low-energy phenomenology in the decoupling limit.

Taking into account dijet or dilepton bounds at the LHC for explaining $B$-meson anomalies with $Z'$,
we consider a minimal extension of the SM with $U(1)'$, that is coupled specifically to heavy flavors and is a linear combination of  $U(1)_{L_\mu-L_\tau}$ and $U(1)_{B_3-L_3}$, as follows,
$$
Q'\equiv y(L_\mu-L_\tau)+x(B_3-L_3)
$$
with $x, y$ being real parameters parametrizing the mixing between $L_\mu-L_\tau$ and $B_3-L_3$.
Here, we note that $B_3-L_3$ is defined for the third generation including one right-handed neutrino and $L_\mu-L_\tau$ are extended to the other two right-handed neutrinos which carry lepton numbers. As $L_\mu$ and $L_\tau$ numbers are assigned to right-handed neutrinos, we get essentially $L_\mu=L_2$ and $L_\tau=L_3$, but we keep $L_\mu-L_\tau$ notations instead of $L_2-L_3$, as motivated by the already known anomaly-free $U(1)'$ symmetries.

As will be discussed in the later section, due to the LHC dimuon constraint on the model, a small value of $x$ is favored phenomenologically.  There are similar $U(1)'$ models based $U(1)_{L_\mu-L_\tau}$ \cite{yavin1} or $U(1)_{B_3-L_3}$ \cite{yanagida} in the literature where extra heavy quarks or lepton mixings are required to explain the $B$-meson anomalies in the former or latter cases. On the other hand, in our case, we assume that both $U(1)_{L_\mu-L_\tau}$ and $U(1)_{B_3-L_3}$ are good symmetries at high energy and only the linear combination of them survives at low energy. Therefore, our setup is qualitatively different from those based on either $U(1)_{L_\mu-L_\tau}$ or $U(1)_{B_3-L_3}$. As will be discussed shortly, this fact becomes manifest in the flavor structure for fermion masses and mixings which require one additional Higgs doublet and extra scalar singlets beyond the SM.
In our model, small $B_3-L_3$ couplings proportional to $x$ might be justified when the third generation fermions get charges under $U(1)'$  through a small gauge kinetic mixing between $L_\mu-L_\tau$ and $B_3-L_3$.  Another origin for small $B_3-L_3$ couplings in our model might be a mismatch between the localization of the $B-L$ gauge boson and the localization of the third generation fermions in the aforementioned gauged $U(1)$ clockwork theory \cite{u1cw}. 

Suppose that we consider a more general combination for $U(1)'$, including anomaly-free  symmetries, $B_1-L_1$ and $B_2-L_2$. Then, the flavor structure for quarks and leptons become more restricted such that we need to introduce more Higgs doublets and complex singlet scalars for realistic quark and lepton masses and mixings. Furthermore, the LHC dilepton bounds constrain the mixing with $B_1-L_1$ and $B_2-L_2$ more strongly than the one with $B_3-L_3$, due to a larger production cross section of $Z'$ via light quarks. Henceforth, we ignore the potential mixing with other generation-dependent $B-L$ symmetries. 

In our model, we introduce two Higgs doublets $H_1, H_2$ and one singlet complex scalars $S$ for spontaneously breaking the electroweak symmetry as well as $U(1)'$, and two right-handed(RH) neutrinos $\nu_{iR}\, (i=2,3)$ for just cancelling the anomalies. Moreover, we also add one more right-handed neutrino $\nu_{1R}$ with zero $U(1)'$ charge for neutrino masses. In order to obtain the consistent neutrino masses and neutrino mixings, we also need to add extra singlet scalars, $\Phi_a (a=1,2,3)$, with $U(1)'$ charges $-y, x+y, x$, respectively.  
The $U(1)'$ charge assignments  are given in Table 1.
\begin{table}[h!]\small
\begin{center}
\begin{tabular}{|c||c|c|c|c|c|c|c|c|c|}
 \hline
 & $q_{3L}$ & $u_{3R}$  &  $d_{3R}$ & $l_{2L}$  & $e_{2R}$  & $\nu_{2R}$ & $l_{3L}$ & $e_{3R}$ & $\nu_{3R}$  \\ [0.5ex]
\hline 
$Q'$ & $\frac{1}{3}x$ & $\frac{1}{3}x$ & $\frac{1}{3}x$ & $y$ & $y$ & $y$ & $-x-y$ & $-x-y$ & $-x-y$ 
\\ [0.5ex]
\hline
\end{tabular}
\\ [0.5ex]
\begin{tabular}{|c||c|c|c|c|c|c|}
 \hline
 &  $S$  & $H_1$ & $H_2$ & $\Phi_1$ & $\Phi_2$ & $\Phi_3$  \\ [0.5ex]
\hline 
$Q'$ &  $-\frac{1}{3}x$ & $0$ & $\frac{1}{3}x$  & $-y$ & $x+y$ & $x$
\\ [0.5ex]
\hline
\end{tabular}
\caption{Nonzero $U(1)'$ charges.}
\label{modelA}
\end{center}
\end{table}

Then, the Lagrangian of our model is the following,
\bea
{\cal L}=-\frac{1}{4}Z'_{\mu\nu} Z^{\prime\mu\nu}+ {\cal L}_S+ {\cal L}_Y
\eea
with
\bea
{\cal L}_S= |D_\mu H_1|^2+|D_\mu H_2|^2+|D_\mu S|^2 - V(H_1,H_2,S)+{\cal L}_{\Phi}
\eea
where the field strength of the $U(1)'$ gauge boson is $Z'_{\mu\nu}=\partial_\mu Z'_\nu -\partial_\nu Z'_\mu$,  the covariant derivatives are $D_\mu \phi_i=(\partial_\mu -ig_{Z'}Q'_{\phi_i} Z'_\mu)\phi_i$ with $\phi_i=H_1,H_2,S$,  $Q'_{\phi_i}$ being $U(1)'$ charge of $\phi_i$, and $g_{Z'}$ being the extra gauge coupling,  and  the scalar potential is given by
\bea
V(H_1,H_2,S)&=& \mu^2_1 |H_1|^2 + \mu^2_2 |H_2|^2-( \mu S H^\dagger_1 H_2+{\rm h.c.}) \nonumber \\
&&+\lambda_1 |H_1|^4+\lambda_2 |H_2|^4 + 2\lambda_3 |H_1|^2|H_2|^2+2\lambda_4 (H^\dagger_1 H_2)(H^\dagger_2 H_1) \nonumber \\
&&+ 2 |S|^2(\kappa_1 |H_1|^2 +\kappa_2  |H_2|^2)+m^2_{S}|S|^2+\lambda_{S}|S|^4.
\eea
Here, ${\cal L}_{\Phi}$ is the scalar Lagrangian due to extra singlet scalars  $\Phi_a(a=1,2,3)$ for lepton flavors, containing the interaction terms between $H_1, H_2$ and $S$. 
Moreover, the Yukawa couplings for quarks and leptons are given by
\bea
-{\cal L}_Y&=&{\bar q}_i (y^u_{ij}H_1+ h^u_{ij}H_2 ) u_j+{\bar q}_i (y^d_{ij} {\tilde H}_1+h^d_{ij} {\tilde H}_2)d_j  \nonumber \\
&&+y^l_{ij} {\bar l}_i {\tilde H}_1 e_j + y^\nu_{ij} {\bar l}_i H_1 \nu_{jR} + \overline {({\nu_{iR}})^c}(M_{ij}+\Phi_a z^{(a)}_{ij})\nu_{jR} +{\rm h.c.}
\eea
with ${\tilde H}_{1,2}\equiv i\sigma_2 H^*_{1,2}$.  
A gauge kinetic mixing between $U(1)_{y(L_\mu-L_\tau)+x(B_3-L_3)}$ and SM hypercharge gauge bosons is also allowed but we assume that it is negligible for our discussion. 

Our model has a rich structure in the Higgs sector which extends the two Higgs doublet models with extra complex singlet scalars. 
As the second Higgs doublet $H_2$ is charged under the local $U(1)'$ symmetry, the Higgs potential has a restricted form.  We note that the Higgs bilinear term is forbidden by $U(1)'$ but it is necessary to get a nonzero pseudo-scalar Higgs mass to be compatible with experiments. Thus, we introduced a complex singlet scalar $S$ with nonzero $U(1)'$ charge. The details of the scalar sector and its phenomenology will be studied elsewhere so we just assume in this work that a correct vacuum with both electroweak symmetry and $U(1)'$ broken exists. After electroweak symmetry and $U(1)'$ are broken spontaneously, the $Z'$ mass is determined to be $m^2_{Z'}=2g^2_{Z'}Q^{\prime 2}_{H_2}\langle H_2\rangle^2+2g^2_{Z'}Q^{\prime 2}_S\langle S\rangle^2+2\sum_{a=1}^3 Q^{\prime 2}_{\Phi_a}\langle\Phi_a\rangle^2$. Thus, as $\langle\Phi_a\rangle$ determines RH neutrino masses, we can take $\langle\Phi_a\rangle \gg \langle H_2\rangle, \langle S\rangle$ such that the $Z'$ mass is larger than weak scale. 

We note that when the second Higgs doublet $H_2$ gets a VEV, it gives rise to a mass mixing between $Z$ and $Z'$ gauge bosons, which is constrained by electroweak precision data as will be discussed in Section 4.5. Henceforth, we assume that the VEV of the second Higgs doublet is small enough such that our model is consistent with electroweak precision data.

As a result, the SM fermion mass matrices in our model are given by the following,
\bea
{\cal L}_Y=- {\bar u} M_u u-{\bar d} M_d d - {\bar l} M_l l - {\bar l} M_D \nu_R -  \overline {({\nu_{R}})^c}M_R \nu_{R}+{\rm h.c.}
\eea
with the flavor structure being
\bea
M_u&=&\left(\begin{array}{ccc} y^u_{11}\langle  H_1\rangle & y^u_{12}\langle H_1\rangle & 0 \\ y^u_{21} \langle H_1\rangle & y^u_{22} \langle H_1 \rangle &  0 \\ h^u_{31} \langle H_2 \rangle & h^u_{32}\langle H_2\rangle &y^u_{33} \langle H_1 \rangle \end{array}\right), \\
M_d&=&\left(\begin{array}{ccc} y^d_{11}\langle {\tilde H}_1\rangle & y^d_{12}\langle {\tilde H}_1\rangle & h^d_{13} \langle {\tilde H}_2\rangle \\ y^d_{21} \langle {\tilde H}_1 \rangle &y^d_{22} \langle {\tilde H}_1\rangle & h^d_{23}\langle {\tilde H}_2\rangle \\ 0 &  0 &y^d_{33}  \langle {\tilde H}_1 \rangle \end{array}\right), \\
M_l &=& \left(\begin{array}{ccc} y^l_{11} \langle {\tilde H}_1 \rangle & 0 & 0 \\ 0 &y^l_{22}  \langle {\tilde H}_1 \rangle & 0 \\ 0 & 0 & y^l_{33} \langle {\tilde H}_1\rangle \end{array}\right), \\
M_D &=& \left(\begin{array}{ccc} y^\nu_{11} \langle H_1 \rangle & 0 & 0 \\ 0 & y^\nu_{22}\langle H_1 \rangle & 0 \\ 0 & 0 &  y^\nu_{33} \langle H_1 \rangle \end{array}\right), \\
M_R &=& \left(\begin{array}{ccc} M_{11} & z^{(1)}_{12} \langle \Phi_1\rangle & z^{(2)}_{13} \langle\Phi_2\rangle \\  z^{(1)}_{21}\langle\Phi_1\rangle & 0 &  z^{(3)}_{23}\langle\Phi_3\rangle \\ z^{(2)}_{31}\langle \Phi_2\rangle & z^{(3)}_{32} \langle\Phi_3\rangle & 0\end{array}\right).      \label{RHmass}
\eea
Then, we find that  the flavor structure is very much restricted due to flavor-dependent $U(1)'$ charges, in particular, the RH neutrino matrix vanishes except  the $(11)$ entry. But, we can generate realistic neutrino masses and mixing after three singlet complex scalars with nonzero $U(1)'$ charges, $\Phi_a (a=1,2,3)$, get nonzero VEVs. Indeed, it has been shown that the above fermion matrices give rise to realistic quark and lepton masses and mixings \cite{flavor,neutrinomix,ko}.  
There are four other categories of neutrino mixing matrices, ${\bf B}_1, {\bf B}_2, {\bf B}_3, {\bf B}_4$ \cite{neutrinomix}, that are compatible with neutrino data in our model. In all those cases, we need at least three complex scalar fields with different $U(1)'$ charges, similarly to the case with (\ref{RHmass}).  It would be interesting to discuss how to distinguish other categories of neutrino mixing matrices in the context of our model.

\section{$B$-meson decays}

We are now in a position to show how to explain the $B$-meson anomalies in our model and identify the relevant parameter space for that.

The Lagrangian for $Z'$ interactions is given by
\bea
{\cal L}_{Z'}&=& g_{Z'} Z'_\mu \Big( \frac{1}{3}x\, {\bar t}\gamma^\mu t+\frac{1}{3}x\, {\bar b}\gamma^\mu b+y {\bar\mu}\gamma^\mu \mu+y\, {\bar \nu}_\mu \gamma^\mu P_L \nu_\mu-(x+y)\,{\bar \tau}\gamma^\mu \tau \nonumber \\
&&-(x+y)\,{\bar \nu}_\tau \gamma^\mu P_L \nu_\tau+y\, {\bar \nu}_{2R}\gamma^\mu P_R \nu_{2R}-(x+y)\,{\bar \nu}_{3R} \gamma^\mu P_R \nu_{3R}\Big). 
\eea
Now we change the basis into the one with mass eigenstates by $d_L=D_L d'_L$ and $u_L=U_L u'_L$ such that $V_{\rm CKM}=U^\dagger_L D_L$.  Taking $U_L=1$ and $D_L=V_{\rm CKM}$, (which is the case for $h^u_{ij}=0$), but no flavor mixing in the charged lepton sector in our model \cite{flavor,muon,ko},
the above $Z'$ interactions become
\bea
{\cal L}_{Z'}&=& g_{Z'} Z'_\mu \Big(\frac{1}{3}x\, {\bar t}'\gamma^\mu t' +\frac{1}{3}x\, {\bar d}'_i \gamma^\mu \Gamma^{dL}_{ij}P_L d'_j+\frac{1}{3}x\, {\bar b}'\gamma^\mu P_R b'  \nonumber \\ 
&&+y {\bar\mu}\gamma^\mu \mu-(x+y)\,{\bar \tau}\gamma^\mu \tau+ y\, {\bar \nu}_\mu \gamma^\mu P_L \nu_\mu-(x+y)\,{\bar \nu}_\tau \gamma^\mu P_L \nu_\tau \nonumber \\
&&+y\, {\bar \nu}_{2R}\gamma^\mu P_R \nu_{2R}-(x+y)\,{\bar \nu}_{3R} \gamma^\mu P_R \nu_{3R}\Big)
\eea
with
\bea
 \Gamma^{dL} &\equiv& V^\dagger_{\rm CKM}\left(\begin{array}{ccc} 0 & 0 & 0 \\ 0 & 0 & 0 \\ 0 & 0 & 1\end{array}\right) V_{\rm CKM} \nonumber \\
 &=&\left(\begin{array}{ccc} |V_{td}|^2 & V^*_{td} V_{ts} & V^*_{td} V_{tb} \\ V^*_{ts} V_{td} & |V_{ts}|^2 & V^*_{ts} V_{tb} \\ V^*_{tb} V_{td} & V^*_{tb} V_{ts} & |V_{tb}|^2 \end{array}\right).
\eea

\begin{figure}[t!]
  \begin{center}
             \includegraphics[height=0.25\textwidth]{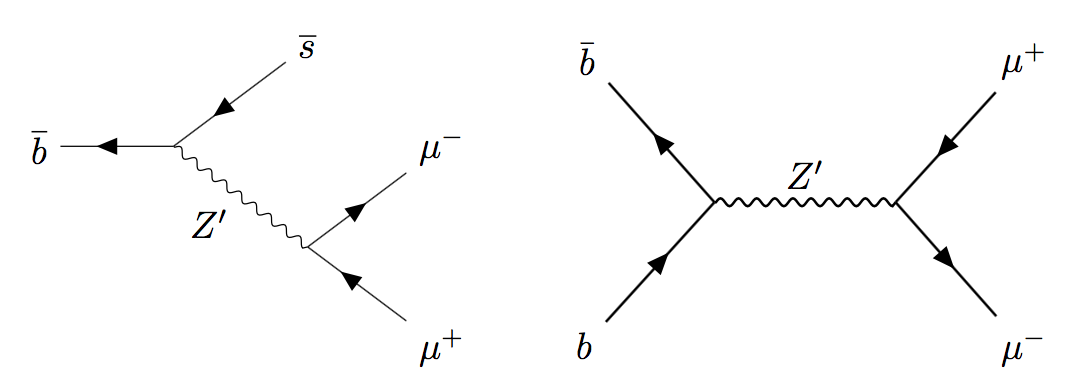}
   \end{center}
  \caption{Feynman diagrams for ${\bar b}\rightarrow {\bar s} \mu^+ \mu^-$ and dimuon production from $Z'$ via $b{\bar b}$ fusion at the LHC. }
  \label{diagrams}
\end{figure}

From the relevant $Z'$ interactions for $B$-meson anomalies and the $Z'$ mass,
\bea
{\cal L}'_{Z'}= g_{Z'} Z'_\mu \Big(\frac{1}{3}x\,V^*_{ts} V_{tb}\,{\bar s}\gamma^\mu P_L b+{\rm h.c.}+y {\bar \mu}\gamma^\mu \mu   \Big)+ \frac{1}{2} m^2_{Z'} Z^{\prime 2}_\mu, \label{zpint}
\eea
we get the classical equation of motion for $Z'$ as
\bea
Z'_\mu= -\frac{g_{Z'}}{m^2_{Z'}} \Big(\frac{1}{3}x\,V^*_{ts} V_{tb}\,{\bar s}\gamma_\mu P_L b+{\rm h.c.}+y {\bar \mu}\gamma_\mu \mu   \Big). \label{zpeq}
\eea
Then, after inserting eq.~(\ref{zpeq}) into eq.~(\ref{zpint}), namely, integrating out the $Z'$ gauge boson, we obtain the  effective four-fermion interaction for ${\bar b}\rightarrow {\bar s}\mu^+ \mu^-$, depicted in Fig.~\ref{diagrams},  as follows,
\bea
{\cal L}_{{\rm eff},{\bar b}\rightarrow {\bar s}\mu^+ \mu^-}= -\frac{xy g^2_{Z'}}{3 m^2_{Z'}}\, V^*_{ts} V_{tb}\,  ({\bar s}\gamma^\mu P_L b) ({\bar \mu}\gamma_\mu \mu)+{\rm h.c.}
\eea
In Fig.~\ref{diagrams},  we also showed one of main production channels for $Z'$ due to $b{\bar b}$ fusion at the LHC. 
Consequently, as compared to the effective Hamiltonian with the SM normalization, 
\bea
\Delta {\cal H}_{{\rm eff},{\bar b}\rightarrow {\bar s}\mu^+ \mu^-} = -\frac{4G_F}{\sqrt{2}}  \,V^*_{ts} V_{tb}\,\frac{\alpha_{em}}{4\pi}\, C^{\mu,{\rm NP}}_9 {\cal O}^\mu_9
\eea
with $ {\cal O}^\mu_9 \equiv ({\bar s}\gamma^\mu P_L b) ({\bar \mu}\gamma_\mu \mu)$ and $\alpha_{\rm em}$ being the electromagnetic coupling, we obtain a new physics contribution to the Wilson coefficient as follows,
\bea
C^{\mu, {\rm NP}}_9= -\frac{8 xy \pi^2\alpha_{Z'}}{3\alpha_{\rm em}}\, \Big(\frac{v}{m_{Z'}}\Big)^2
\eea
with $\alpha_{Z'}\equiv g^2_{Z'}/(4\pi)$,
and vanishing contributions for other operators \cite{otherops}, $C^{\mu,{\rm NP}}_{10}=C^{\prime\mu, {\rm NP}}_9=C^{\prime\mu, {\rm NP}}_{10}=0$.
We note that $xy>0$ is chosen for a negative sign of $C^\mu_9$, being consistent with $B$-meson anomalies.
Therefore, requiring the best-fit value, $C^{\mu, {\rm NP}}_9=-1.10$ \cite{muon}, (while taking $[-1.27,-0.92]$ and $[-1.43,-0.74]$ within $1\sigma$ and $2\sigma$ errors), to explain the $B$-meson anomalies, we need 
\be
m_{Z'}=\Big(xy\, \frac{\alpha_{Z'}}{\alpha_{\rm em}} \Big)^{1/2}\, 1.2\,{\rm TeV}.  
\ee
Therefore, we need the $Z'$ mass of about $1\,{\rm TeV}$ for $xy\sim 1$ and $\alpha_{Z'}\sim \alpha_{\rm em}$.
However, the $Z'$ mass can be smaller for values of $xy$ less than unity or $\alpha_{Z'}\lesssim \alpha_{\rm em}$. In next section, we will discuss other phenomenological constraints on these parameters.

\section{Constraints on the model}

In this section, we consider the constraints on the model from other meson decays such as $B_s-{\bar B}_s$ mixing, anomalous magnetic moment of muon, tau decays, neutrino trident production, electroweak precision test, and dimuon resonance searches at the LHC.

\subsection{Other meson processes}

Our model predicts a new contribution to the effective four-fermion interactions for other meson decays and meson mixings.

First, the most stringent constraints stem from the $B^0_i-{\bar B}^0_i$ mixings with $i=s,d$ that receive new contributions due to $Z'$ interactions as follows, 
\bea
{\cal L}_{{\rm eff},B^0_i-{\bar B}^0_i}=-\frac{g^2_{Z'}}{2m^2_{Z'}}\, \Big(\frac{1}{3}xV^*_{ti} V_{tb} \Big)^2 ({\bar d}_i\gamma^\mu P_L b)({\bar d}_i\gamma_\mu P_L b).
\eea
As a result, we get a correction to the effective Hamiltonian with the SM normalization, as follows,
\bea
\Delta {\cal H}_{{\rm eff},B^0_i-{\bar B}^0_i}= \frac{G^2_F m^2_W}{16\pi^2}\, (V^*_{ti} V_{tb})^2 \, C^{i,{\rm NP}}_{VLL}\,  ({\bar d}_i\gamma^\mu P_L b)({\bar d}_i\gamma_\mu P_L b)
\eea
with
\bea
C^{i,{\rm NP}}_{VLL}&=&\frac{16\pi^2 x^2}{9} \, \frac{g^2_{Z'} v^4}{m^2_{Z'} m^2_W}  \nonumber \\
&=& 0.39 \Big(\frac{x}{0.06}\Big)^2g^2_{Z'} \Big(\frac{300\,{\rm GeV}}{m_{Z'}}\Big)^2. \label{bbar}
\eea
Then, the mass difference in the $B_s$ system is given by
\bea
\Delta M_i = \frac{2}{3} m_{B_i} f^2_{B_i} {\hat B}_{B_i}  \frac{G^2_F m^2_W}{16\pi^2}\, (V^*_{ti} V_{tb})^2 \,\Big(C^{i,{\rm SM}}_{VLL}+C^{i,{\rm NP}}_{VLL}\Big).
\eea
For instance, $C^{s,{\rm SM}}_{VLL}\simeq 4.95$ \cite{MsSM}. 
Experimental and SM values of $\Delta M_s$ are given by
 $(\Delta M_s)^{\rm SM}=(17.4\pm 2.6)\,{\rm ps}^{-1}$ \cite{MsSM} and $(\Delta M_s)^{\rm exp}=(17.757\pm 0.021)\,{\rm ps}^{-1}$  \cite{Msexp}, respectively. 
Therefore, the new physics contribution for $B^0_s-{\bar B}^0_s$ mixing is constrained by $C^{s,{\rm NP}}_{VLL}\lesssim 0.15(0.30)\times C^{s,{\rm SM}}_{VLL}\simeq 0.74(1.48)$ within $1\sigma(2\sigma)$ errors of the SM value.  Moreover, the $B^0_d-{\bar B}^0_d$ or $K^0-{\bar K}^0$ mixings are similarly modified by the $Z'$ interactions, but they constrain the model less than in the case of $B^0_s-{\bar B}^0_s$ mixing due to larger experimental and theoretical errors \cite{flavor}.

In our model, we can choose a small value of $x$ (i.e. small $Z'$ couplings to quarks) to satisfy the dimuon bounds from the LHC as will be discussed in next section. Then, in view of the Wilson coefficient given in eq.~(\ref{bbar}), the bounds from the $B^0_s-{\bar B}^0_s$, $B^0_d-{\bar B}^0_d$ as well as $K^0-{\bar K}^0$ mixings are easily satisfied in all the parameter space where the $B$-meson anomalies and the LHC dimuon bound are accommodated.  In Fig.~\ref{Combine2}, in the parameter space for $m_{Z'}$ and $g_{Z'} x$ with $g_{Z'} y=1, 1.5$, the red region is excluded by the bound from the  $B^0_s-{\bar B}^0_s$ within $1\sigma$ of the SM errors, but it is already well constrained by the other bounds. 

We also note that for minimal flavor violation scenario as in our model,  the new contribution in $C^\mu_9$ for $B$-meson decays has a correlation with those for rare kaon decays, modifying the decay rate of kaon, ${\cal B}(K^+\rightarrow \pi^+\nu{\bar\nu})$, but the effect in $K^+\rightarrow \pi^+\nu{\bar\nu}$ is small as compared to experimental uncertainties \cite{kaon}. 
On the other hand, there is no bound on $Z'$ interactions due to $B_{s,d}\rightarrow \mu^+\mu^-$ or $K_L\rightarrow \mu^+\mu^-$ in our model \cite{crivellin}, as the $Z'$ couplings to muons are vector-like.
However, those rare decay modes could constrain flavor-changing interactions of heavy Higgs bosons in our model, although they can be evaded for a small VEV of the second Higgs doublet \cite{crivellin}.  

Finally, we also comment on inclusive radiative decays such as $B \to X_s\gamma$ that could give rise to important constraints on flavor-violating interactions from new physics. 
The effective Hamiltonian relevant for $b\rightarrow s\gamma$ transition is
\bea
{\cal H}_{{\rm eff},b\rightarrow s\gamma }= -\frac{4G_F}{\sqrt{2}} \, V_{tb} V^*_{ts} \Big(C_7 {\cal O}_7+ C_8 {\cal O}_8 \Big)
\eea
with
\bea
{\cal O}_7 &=& \frac{e}{16\pi^2}\, m_b\, {\bar s} \sigma^{\mu\nu} P_R b\, F_{\mu\nu}, \\
{\cal O}_8 &=& \frac{g_S}{16\pi^2}\, m_b\,  {\bar s} \sigma^{\mu\nu} P_R  T^a b\,  G^a_{\mu\nu}.
\eea
Then, the $Z'$ contributions to the Wilson coefficients \cite{buras} are given by
\bea
C^{\rm BSM}_7=-\frac{1}{324} \frac{x^2 g^2_{Z'} v^2}{m^2_{Z'}}=-\frac{1}{3} C^{\rm BSM}_8.  \label{C7}
\eea
The NNLO SM prediction for ${\cal B}(B\rightarrow X_s \gamma)$ \cite{bsg-th} is 
\be
{\cal B}(B\rightarrow X_s \gamma) = (3.36\pm 0.23)\times 10^{-4}
\ee
whereas the experimentally measured value of ${\cal B}(B\rightarrow X_s \gamma)$ from HFAG \cite{bsg-exp} is
\be
{\cal B}(B\rightarrow X_s \gamma) =(3.43\pm 0.21\pm 0.07)\times 10^{-4}. 
\ee
As a result, the SM prediction for $B\rightarrow X_s \gamma$ is consistent with experiments, so we obtain the bounds on the modified Wilson coefficients as $|C^{\rm BSM}_7/C^{\rm SM}_7|\lesssim 0.08 (0.15)$ at $1\sigma (2\sigma)$ level.  Consequently, from eq.~(\ref{C7}),  we get the bound on $Z'$ mass in our model as $m_{Z'}/(x g_{Z'})\gtrsim 81(59)\,{\rm GeV}$, which is well satisfied in the parameter space where we can explain the $B$-meson anomalies.

\subsection{Anomalous magnetic moments of muon}

Interactions of $Z'$ gauge boson to leptons lead to corrections to the anomalous magnetic moments of leptons as follows \cite{ael},
\bea
a^{Z'}_l= \frac{Q^{\prime 2}_l \alpha_{Z'}}{2\pi}\, F\Big(\frac{m_{Z'}}{m_l} \Big)
\eea 
where $Q^{\prime }_l $ is the $U(1)'$ charge of lepton $l$ and the loop function is given by
\bea
F(x)\equiv \int_0^1 dz\, \frac{2z(1-z)^2}{(1-z)^2+x^2 z}. 
\eea
For $m_{Z'}\gg m_l$, the anomalous magnetic moment becomes
\bea
a^{Z'}_l\approx \frac{Q^{\prime 2}_l \alpha_{Z'}}{3\pi}\, \frac{m^2_l}{m^2_{Z'}}. 
\eea
Therefore, the anomalous magnetic moment  of muon becomes
\bea
a^{Z'}_\mu \approx  151\times 10^{-11}\Big(\frac{y g_{Z'}}{2} \Big)^2 \Big(\frac{500\,{\rm GeV}}{m_{Z'}} \Big)^2
\eea 
The deviation of the anomalous magnetic moment of muon between experiment and SM values is given \cite{amu,pdg} by
\bea
\Delta a_\mu = a^{\rm exp}_\mu - a^{\rm SM}_\mu=288(80)\times 10^{-11},
\eea
which is a $3.6\sigma$ discrepancy from the SM \cite{pdg}.

As shown in Fig.~\ref{Combine1},  with a small $x$ (i.e. small $Z'$ couplings to quarks), there is a viable parameter space of $m_{Z'}$ and $g_{Z'}$, explaining the $B$-meson anomalies as well as $(g-2)_\mu$ within $2\sigma$ simultaneously, but the $(g-2)_\mu$ region is not consistent with bounds from tau decays and neutrino trident production, as will be discussed in next subsection.

\subsection{Tau decays}

One-loop box diagram containing the $Z'$ gauge boson in our model leads to corrections to tau decay processes. Thus, the decay rate for $\tau\rightarrow \mu\nu_\tau {\bar\nu}_\mu$ is encoded in the following ratio of branching ratios of tau decay \cite{yavin1},
\bea
\frac{{\rm BR}(\tau\rightarrow \mu \nu_\tau {\bar\nu}_\mu)}{{\rm BR}(\tau\rightarrow \mu \nu_\tau {\bar\nu}_\mu)_{\rm SM}}=1+\Delta
\eea
with
\bea
\Delta= \frac{3 y(x+y) g^2_{Z'}}{4\pi^2}\frac{{\rm log}(m^2_W/m^2_{Z'})}{1-m^2_{Z'}/m^2_W}. 
\eea
The PDG value \cite{pdg} for ${\rm BR}(\tau\rightarrow \mu \nu_\tau {\bar\nu}_\mu)$ reads
\bea
{\rm BR}(\tau\rightarrow \mu \nu_\tau {\bar\nu}_\mu)_{\rm exp}= (17.39\pm 0.04)\%,
\eea 
while the SM prediction for the branching ratio with the tau lifetime, $\tau_\tau=(290.29\pm 0.53)\times 10^{-15}\,{\rm s}$, taken into account \cite{yavin1}, is
\bea
{\rm BR}(\tau\rightarrow \mu \nu_\tau {\bar\nu}_\mu)_{\rm SM}=(17.29\pm 0.03)\%.
\eea
Therefore, there is an excess with more than $2\sigma$ in ${\rm BR}(\tau\rightarrow \mu \nu_\tau {\bar\nu}_\mu)$ but we can use the above result to constrain $\Delta$  as
\bea
\Delta =(5.8\pm 3.0)\times 10^{-3}.
\eea

\subsection{Neutrino trident production}

In the presence of a heavy $Z'$, there appears a leptonic 4-fermion operator of type,
\be
 {\cal L}_{{\rm eff},\nu_\mu\rightarrow \nu_\mu\mu{\bar\mu}}=-\frac{y^2 g^2_{Z'}}{m^2_{Z'}}\,({\bar\mu}\gamma^\mu \mu)({\bar\nu}_\mu\gamma_\mu P_L\nu_\mu).
 \ee 
Then, the above effective operator modifies the total cross section of neutrino trident production, $\nu_\mu+N\rightarrow \nu_\mu+N+\mu{\bar \mu}$, as follows \cite{yavin1,yavin2},
\bea
\frac{\sigma}{\sigma_{\rm SM}}\simeq \frac{1+(1+4s^2_W+2 y^2 g^2_{Z'} v^2 /m^2_{Z'})^2}{1+(1+4s^2_W)^2}. 
\eea
Combining the SM predictions with the measured cross sections of neutrino trident production from CHARM-II \cite{charm}, CCFR\cite {ccfr} and NuTeV \cite{nutev1} (later update \cite{nutev2})  and taking the weighted average \cite{yavin1}, we get
\bea
\frac{\sigma_{\rm exp}}{\sigma_{\rm SM}}= 0.83(0.95)\pm 0.18(0.25). 
\eea
As a consequence, we get the constraint on $Z'$ at the $2\sigma$ level as follows,
\bea
\frac{m_{Z'}}{y g_{Z'}} >820(554)\,{\rm GeV}.
\eea

\subsection{Electroweak precision data}

In this section, we comment how electroweak precision tests our model. 
Since the second Higgs $H_2$ carries a nonzero $B_3-L_3$ charge in our model, the VEV of the second Higgs doublet is constrained by electroweak precision data, due to the mass mixing between $Z$ and $Z'$ gauge bosons at tree level. 
In our model, the $Z$-boson mass gets modified to
\bea
m^2_Z=\frac{1}{2}\left[m^2_{Z,0}+m^2_{Z'}-\sqrt{(m^2_{Z'}-m^2_{Z,0})^2+s^{-2}_W c^{-2}_W e^2 g^2_{Z'} Q^{\prime 2}_{H_2} v^4_2} \right]
\eea
where $s_W\equiv \sin\theta_W, c_W\equiv \cos\theta_W$,  $m^2_{Z,0}\equiv \frac{1}{4} (g^2+g^2_Y) v^2$ with $v=\sqrt{v^2_1+v^2_2}=246\,{\rm GeV}$, $v_{1,2}\equiv \sqrt{2}\langle H_{1,2}\rangle$, and $m^2_{Z'}$ is the squared $Z'$ mass, obtained from the VEVs of scalar fields with nonzero $U(1)'$ in our model.
Moreover, the mixing angle between $Z$ and $Z'$ gauge bosons is given by
\bea
\tan(2\zeta)= -\frac{s^{-1}_W c^{-1}_W e g_{Z'} Q'_{H_2} v^2_2}{m^2_{Z'}-m^2_{Z}}.
\eea
Thus, the resulting correction to the $\rho$ parameter in our model is given by
\bea
\Delta\rho&=&\frac{m^2_W}{m^2_{Z}\cos^2\theta_W}-1 \nonumber \\
&\sim & 10^{-4} \left(\frac{x}{0.1}\right)^2 g^2_{Z'} \left(\frac{v_2}{v/2} \right)^4 \left(\frac{200\,{\rm GeV}}{m_{Z'}}\right)^2.
\eea
The PDG value reads $\Delta\rho=(3.7\pm 2.3)\times 10^{-4}$ \cite{pdg}, so our model is consistent with electroweak precision test, as far as the VEV of the second Higgs and/or the mixing with $B_3-L_3$ in $U(1)'$ are small enough and/or $Z'$ is heavy enough \footnote{We would like to thank Yuji Omura for pointing out this point.}.  As the VEV of the second Higgs doublet is model-dependent, namely, it depends on whether $h^d_{13}$ and $h^d_{23}$ in the down-type Yukawa matrix, are small or not, we postpone the detailed discussion on phenomenology in the Higgs sector to a future work.

\subsection{Dimuon resonance searches at the LHC}

Suppose that  right-handed neutrinos,  extra Higgs and singlet scalars are heavier than $Z'$. 
Then, $Z'$ decays into SM particles only, namely, $t{\bar t}$, $b{\bar b}$, $\mu{\bar \mu}$,  $\tau{\bar \tau}$, and $\nu_{\mu,\tau}{\bar\nu}_{\mu,\tau}$, while the decay modes into $b{\bar s}, {\bar s}b$ or other down-type quarks are suppressed by CKM mixings.  But, if right-handed neutrinos or extra scalars can be light enough, there can be other $Z'$ decay modes, leading to potentially interesting signatures for $Z'$ searches.

\begin{figure}[t!]
  \begin{center}
        \includegraphics[height=0.40\textwidth]{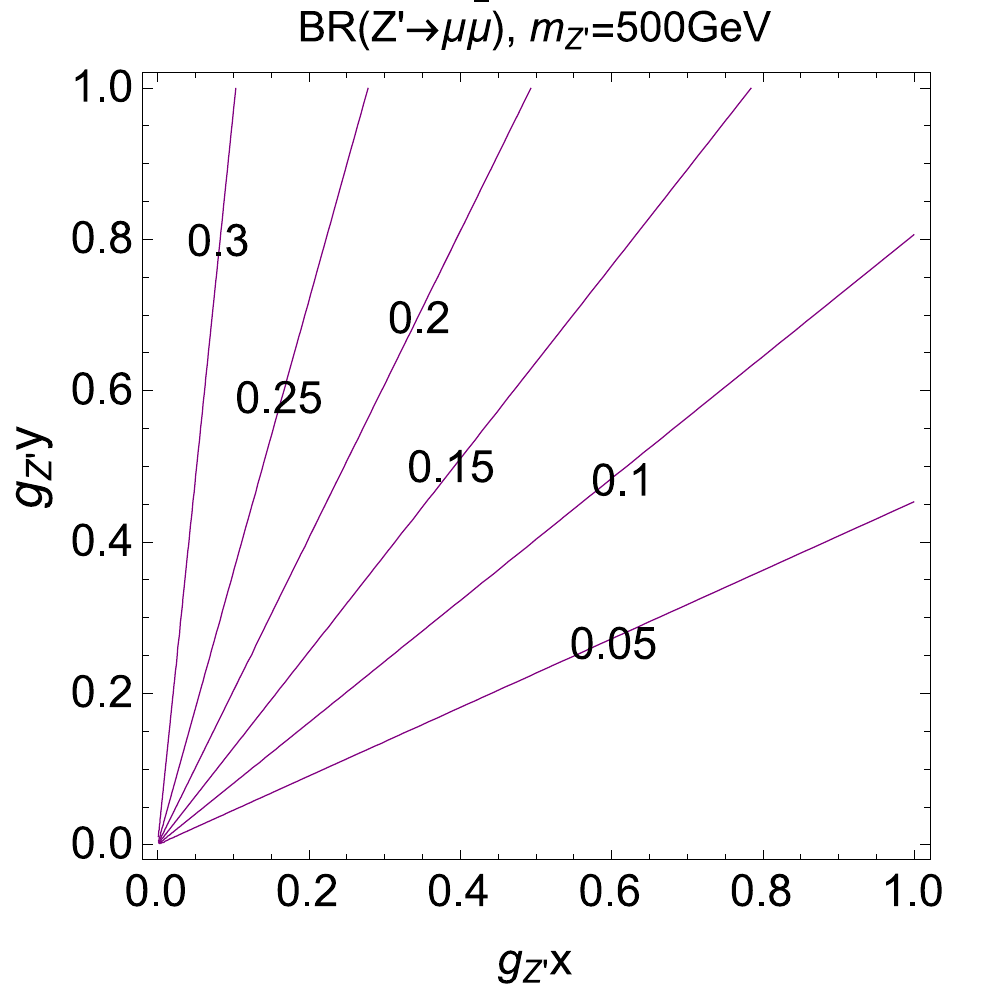}
      \includegraphics[height=0.40\textwidth]{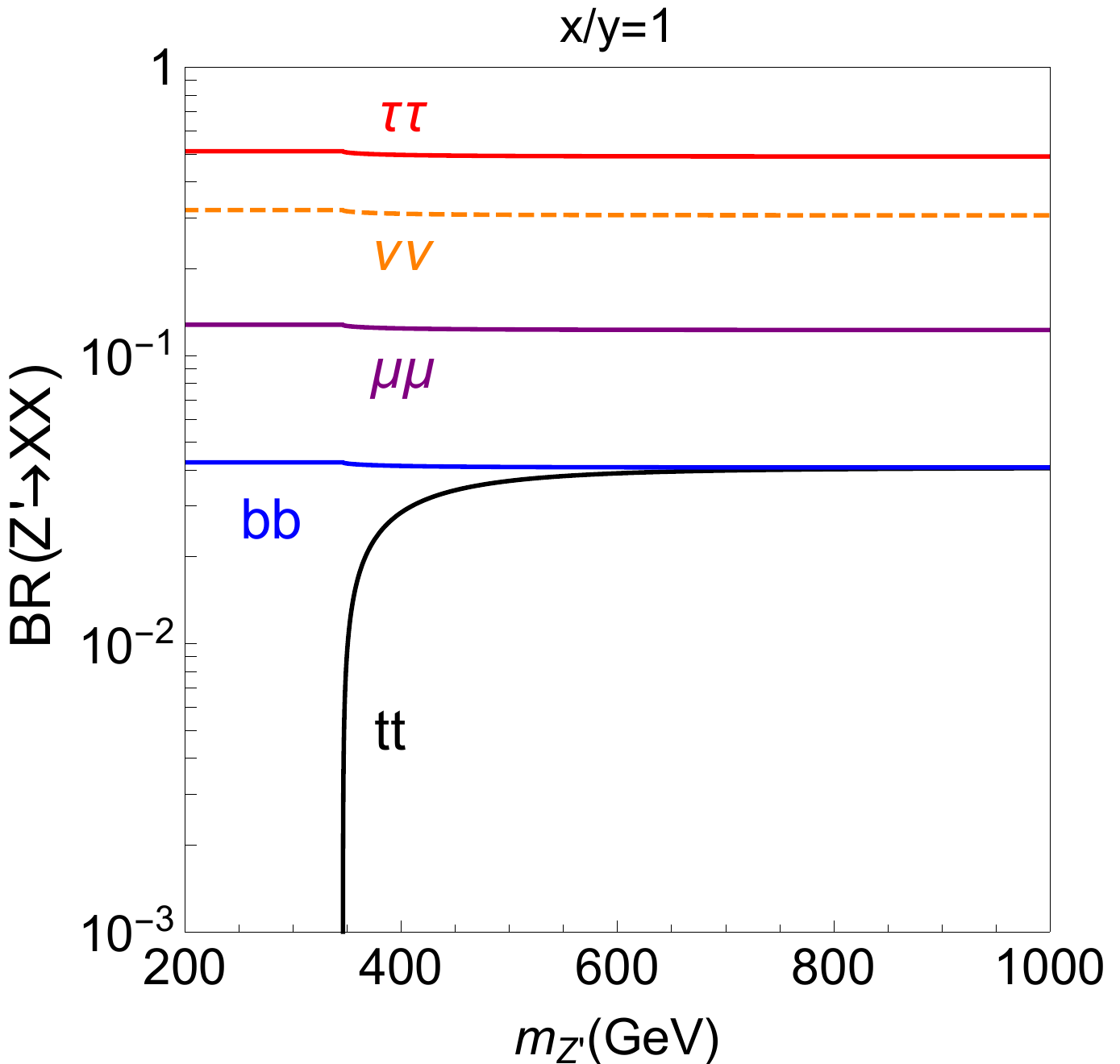} \\
      \includegraphics[height=0.40\textwidth]{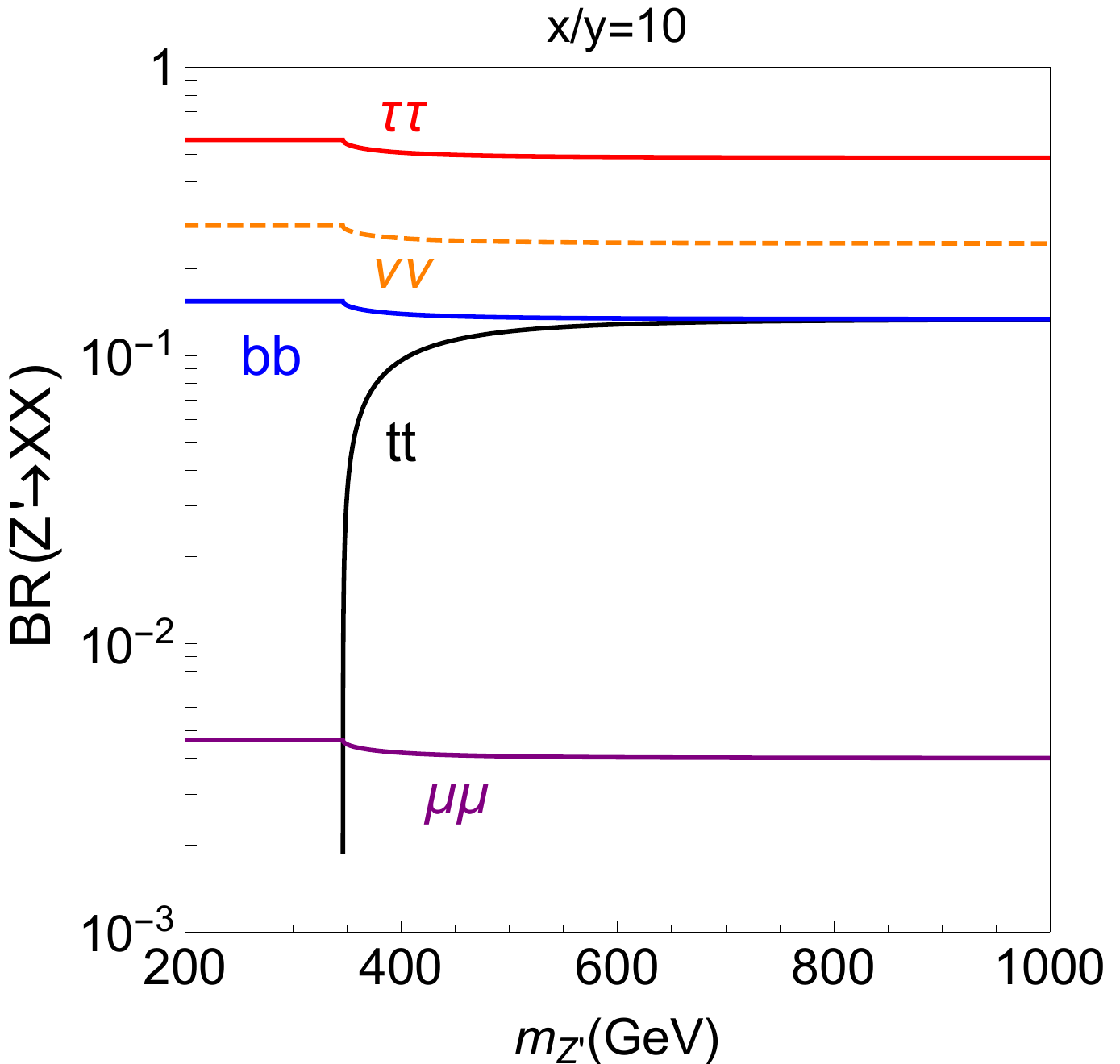}
      \includegraphics[height=0.40\textwidth]{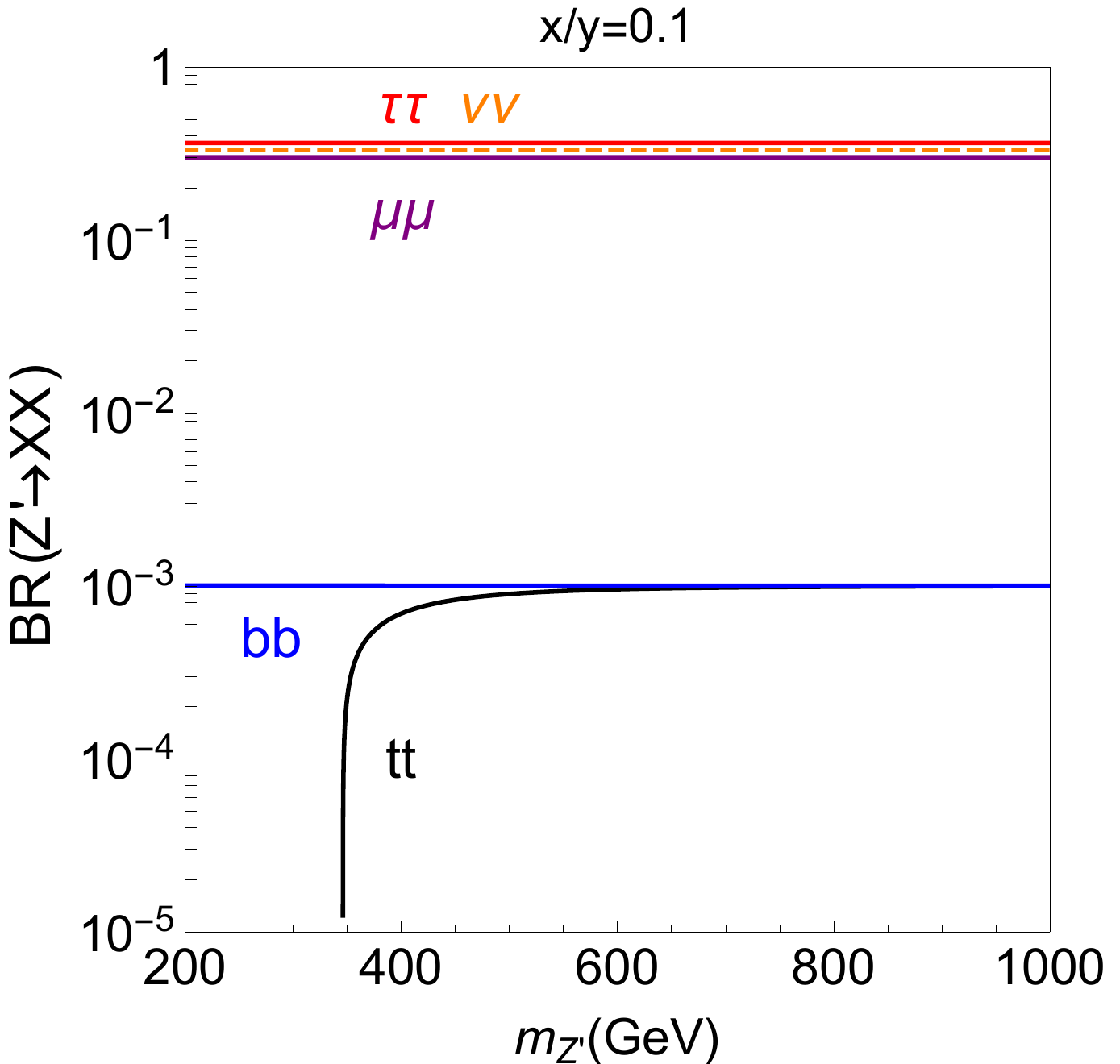}
   \end{center}
  \caption{Upper left: Contours of ${\rm BR}(Z'\rightarrow \mu {\bar \mu})$ in the parameter space of $g_{Z'} x$ vs $g_{Z'} y$. Rest: Branching ratios of $Z'$ decay rates. We have taken $x/y=1, 0.1, 10$ clockwise. }
  \label{BRmz}
\end{figure}

The partial decay rates of $Z'$ are given as follows,
\bea
\Gamma(Z'\rightarrow t{\bar t},b{\bar b}) &=&\frac{x^2 g^2_{Z'}}{36\pi}\, m_{Z'}\left(1+\frac{2m^2_{t,b}}{m^2_{Z'}} \right)\sqrt{1-\frac{4m^2_{t,b}}{m^2_{Z'}}} \,,\\
\Gamma(Z'\rightarrow \mu {\bar \mu}) &=& \frac{y^2 g^2_{Z'}}{12\pi}\, m_{Z'}\left(1+\frac{2m^2_\mu }{m^2_{Z'}} \right) \sqrt{1-\frac{4m^2_\mu}{m^2_{Z'}}}\,, \\
\Gamma(Z'\rightarrow \tau {\bar \tau}) &=& \frac{(x+y)^2 g^2_{Z'}}{12\pi}\, m_{Z'} \left(1+\frac{2m^2_\tau }{m^2_{Z'}} \right) \sqrt{1-\frac{4m^2_\tau}{m^2_{Z'}}}\,, \\
 \Gamma(Z'\rightarrow \nu_\mu {\bar \nu}_\mu) &=& \frac{ y^2g^2_{Z'}}{24\pi}\, m_{Z'},  \\
 \Gamma(Z'\rightarrow \nu_\tau {\bar \nu}_\tau) &=& \frac{(x+y)^2 g^2_{Z'}}{24\pi}\, m_{Z'}
\eea
where we ignored the masses for light fermions. 
Then, in the limit of heavy $Z'$, the decay branching fractions are ${\rm BR}(t{\bar t},b{\bar b}):{\rm BR}(\mu{\bar\mu}):{\rm BR}(\tau{\bar\tau)}: {\rm BR}(\nu_\mu{\bar\nu}_\mu+\nu_\tau {\bar\nu}_\tau)=2x^2:6y^2:6(x+y)^2:3(y^2+(x+y)^2)$.
Thus, for $x\ll y$, $Z'$ decays mostly into $\mu{\bar \mu},\tau{\bar\tau}, \nu_\mu {\bar\nu}_\mu, \nu_\tau {\bar\nu}_\tau$ with ${\rm BR}(\mu{\bar \mu},\tau{\bar\tau}):{\rm BR}(\nu_\mu {\bar\nu}_\mu+ \nu_\tau {\bar\nu}_\tau)\sim1:1$, as in $U(1)_{L_\mu-L_\tau}$ models.
In the upper left plot of Fig.~\ref{BRmz}, ${\rm BR}(Z'\rightarrow \mu{\bar\mu})$ is shown in the parameter space of $g_{Z'} x$ vs $g_{Z'} y$. As $x$ gets larger(smaller) than $y$, the branching fraction of dimuon channel becomes smaller (larger).  In the rest plots of Fig.~\ref{BRmz}, we also show the branching ratios for all the allowed decay modes depending on  the ratio $x/y$. 
For all values of $x/y$, the $\tau{\bar\tau}$ mode is dominant. For $x\lesssim y$, the $\mu{\bar\mu}$ mode is sizable up to about $30\%$ while $b{\bar b}, t{\bar t}$ modes are sub-dominant. But, for $x\gg y$, the $b{\bar b}, t{\bar t}$ modes are sizable while the $\mu{\bar\mu}$ mode is negligible about $0.5\%$.

\begin{figure}[t!]
  \begin{center}
      \includegraphics[height=0.50\textwidth]{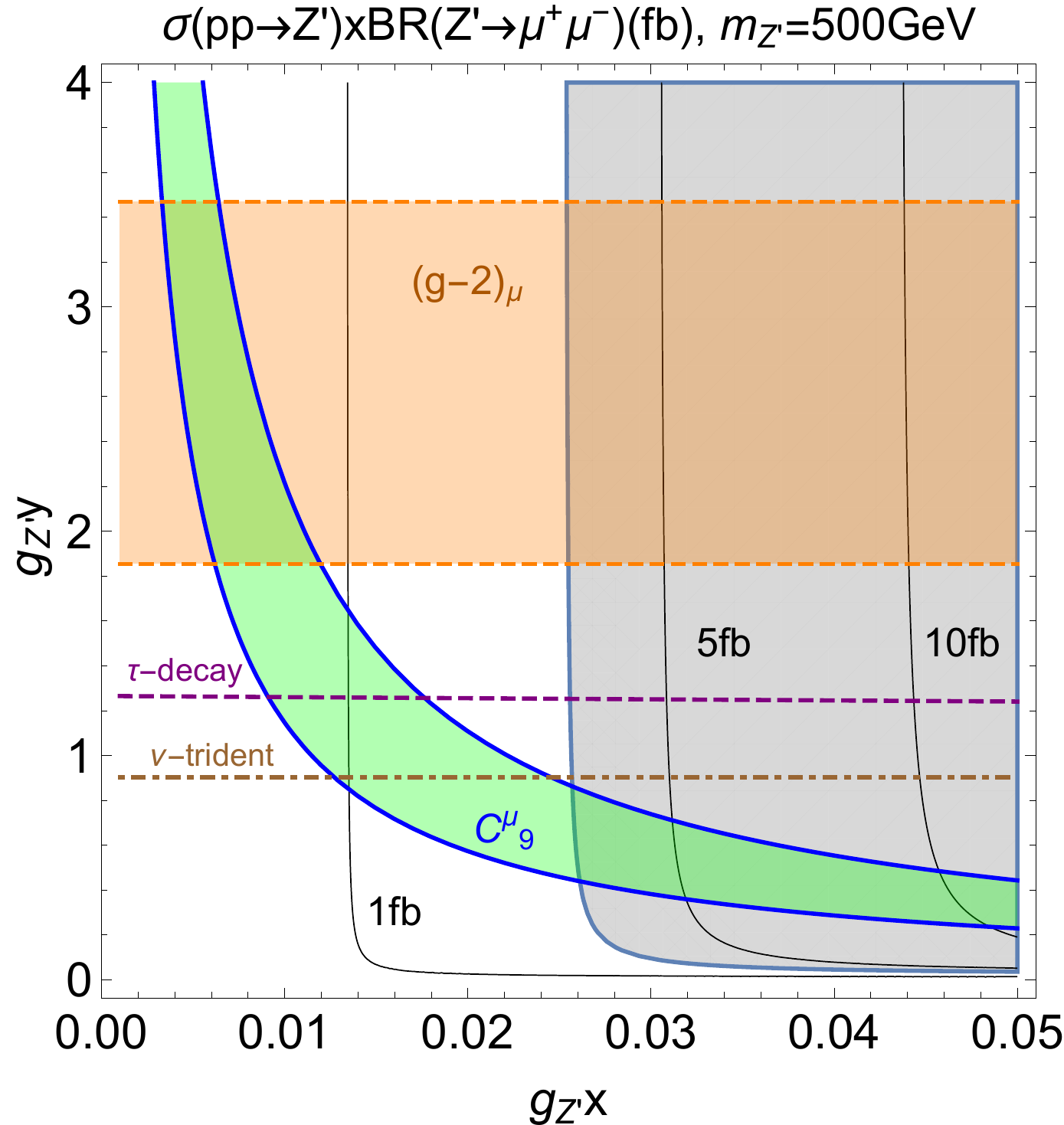}
         \includegraphics[height=0.50\textwidth]{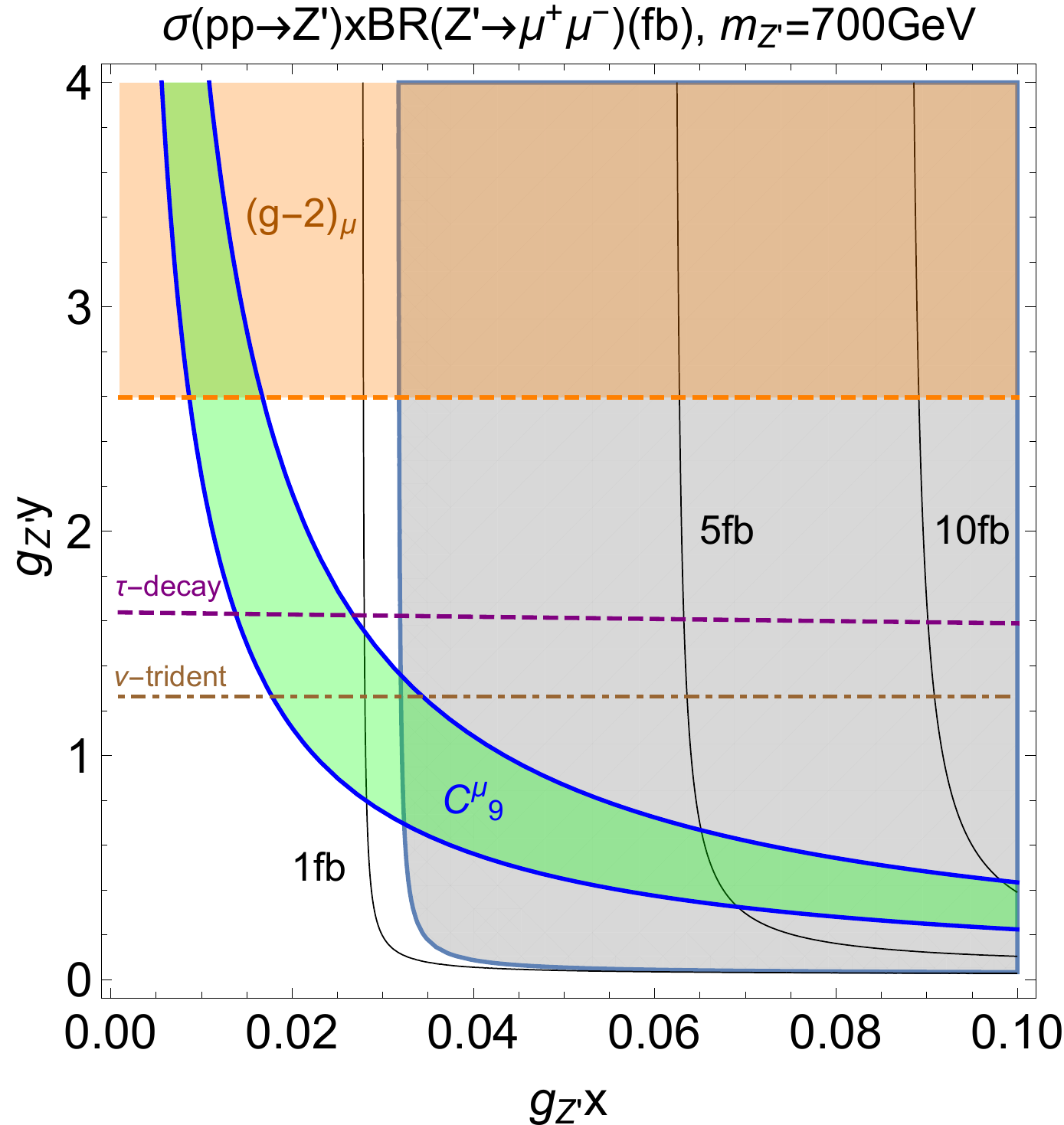}
   \end{center}
  \caption{Combined plots in the parameter space for $g_{Z'}x$ vs $g_{Z'} y$.  The regions favored by $B$-meson anomalies and $(g-2)_\mu$  at $2\sigma$ are shown in green and orange, respectively, and the region excluded by dimuon $Z'$ searches is shown in gray. The regions above purple dashed and brown dot-dashed lines are excluded by $\tau\rightarrow \mu{\bar \nu}_\mu \nu_\tau$ and neutrino trident production, respectively. 
  We have taken $m_{Z'}=500(700)\,{\rm GeV}$ on left (right). Contours for $\sigma(pp\rightarrow Z')\times {\rm BR}(Z'\rightarrow \mu{\bar\mu})=1, 5, 10 (1,5,10,20,30)$ fb are also shown on left (right).}
  \label{Combine1}
\end{figure}

Now we turn to the discussion on the $Z'$ production channels and the LHC dimuon constraints on our model.
To investigate the LHC constraints on the dimuon decay modes of $Z'$, we used the most recent results coming from ATLAS~\cite{ATLAS:2017wce}, while there are dimuon searches from CMS with similarly strong bounds too \cite{Khachatryan:2016zqb}.
For the calculation of the production cross section, we used MadGraph5\_aMC@NLO~\cite{Alwall:2014hca} with the model file being implemented in
FeynRules~2.0~\cite{Alloul:2013bka}. Here we adopted the PDF set of NN23LO1 but did not take into account the $K$-factor.
There are three dominant channels for $Z'$ production at the LHC: $pp\rightarrow Z'$ through the $b\bar{b}$ fusion as shown in the right Feynman diagram in Fig.~\ref{diagrams} (while $b\bar{s}$ fusion is sub-dominant), $pp\rightarrow Z' j$, and $pp\rightarrow Z' jj$. The contribution of the subdominant channel, $pp\rightarrow Z' t\bar{t}$, is around $O(10^{-2})$ of those three channels, and the top quark box diagram contributions for $pp\rightarrow Z'Z(g)$ are negligible too.
The rough ratios of $Z'$ production cross sections are $\sigma(b{\bar b}\rightarrow Z'): \sigma(gg\rightarrow Z'g):\sigma(gg\rightarrow Z'Z)\sim100:1:10^{-2}$.  
The production rate of $Z'$ is characterized by $g_{Z'}$ and the parameter $x$ parametrizing the $(B_3-L_3)$ charge while ${\rm  BR}(Z'\rightarrow \mu{\bar\mu})$ is determined by the ratio of $(L_\mu-L_\tau)$ and $(B_3-L_3)$ charges, as shown in the upper left plot of Fig.~\ref{BRmz}.

\begin{figure}[t!]
  \begin{center}
      \includegraphics[height=0.47\textwidth]{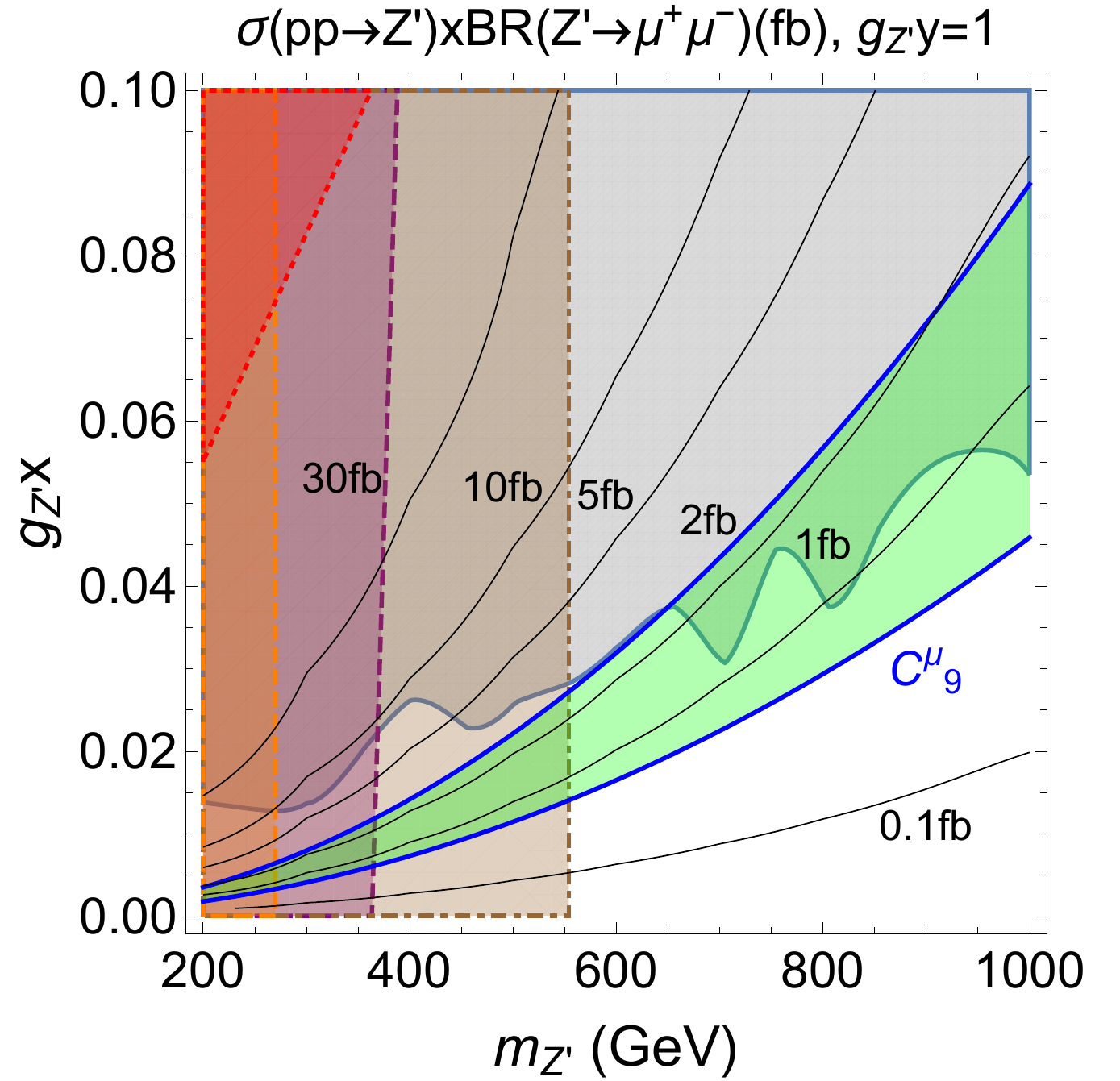}
      \includegraphics[height=0.47\textwidth]{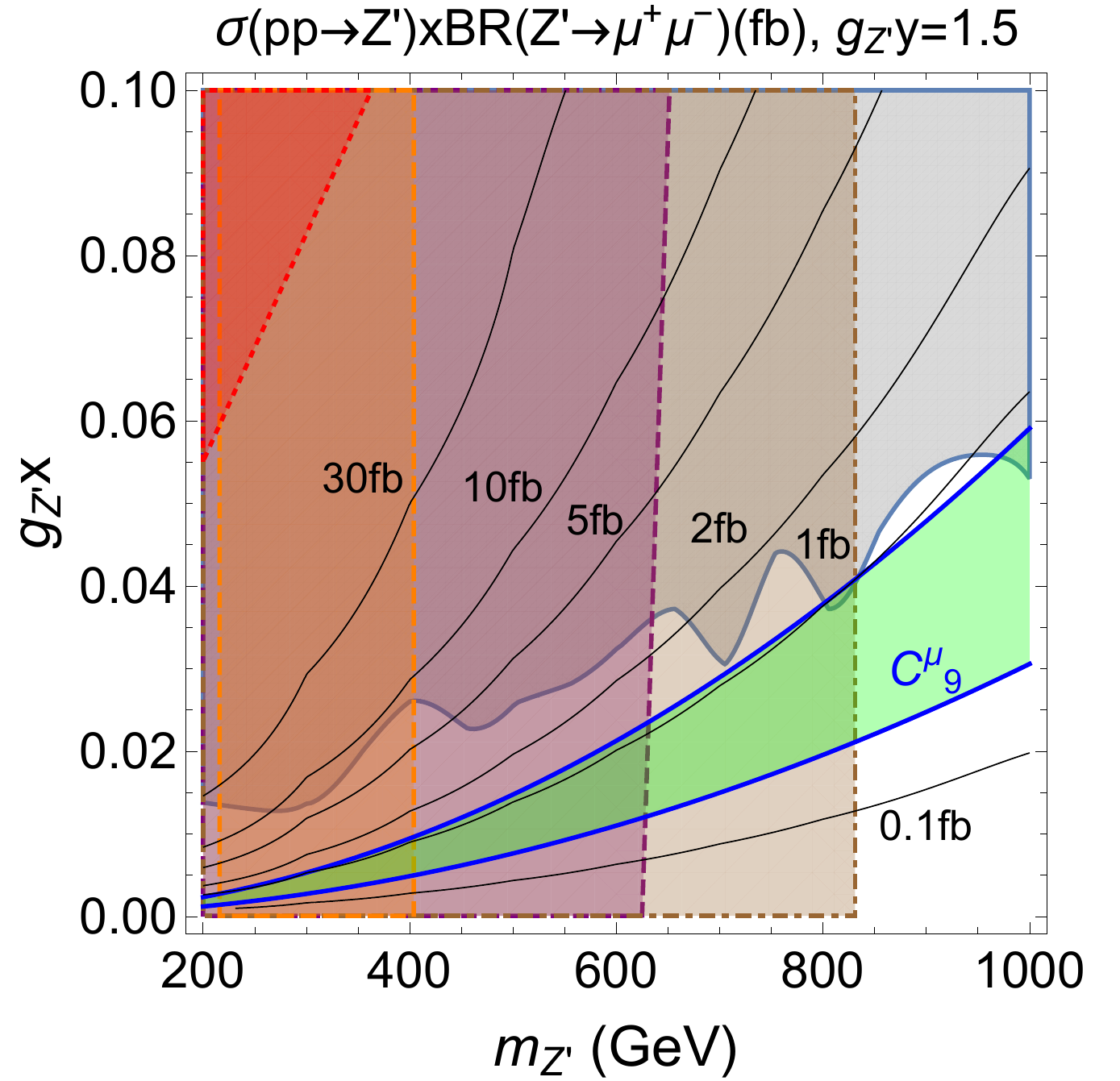}
   \end{center}
  \caption{Combined plot  in the parameter space for $m_{Z'}$ vs $g_{Z'} x$. The regions favored by $B$-meson anomalies and $(g-2)_\mu$  at $2\sigma$ are shown in green and orange, respectively, and the region excluded by dimuon $Z'$ searches is shown in gray. The regions excluded by $\tau\rightarrow \mu{\bar \nu}_\mu \nu_\tau$ and neutrino trident production are shown in purple and brown, respectively.  The red regions are also excluded by $B_s-{\bar B}_s$ mixing within $1\sigma$ for the SM errors. We have taken $g_{Z'} y=1(1.5)$ on left (right). Contours for $\sigma(pp\rightarrow Z')\times {\rm BR}(Z'\rightarrow \mu^+\mu^-)=1, 2, 5, 10, 30$ fb are also shown in both plots.}
  \label{Combine2}
\end{figure}

 In Fig.~\ref{Combine1}, for $m_{Z'}=500, 700$ GeV as benchmark points,  we showed the region excluded by the ATLAS bounds (in gray) in the parameter space of $g_{Z'}x$ vs $g_{Z'} y$, and depicted the regions favored  by $B$-meson anomalies (in green) and $(g-2)_\mu$ (in orange). 
We find that  $(g-2)_\mu$ could not be simultaneously accommodated with $B$-meson anomalies, due to the bounds from $\tau$ decays and neutrino trident production. The LHC dimuon searches constrain the parameter space to the region with a sizable $x$. 

In Fig.~\ref{Combine2}, varying $Z'$ mass and $g_{Z'} x$ for a fixed value, $g_{Z'} y=1$ (left) or $g_{Z'} y=1.5$ (right), we showed the interplay between the LHC bounds and $(g-2)_\mu$ to impose a tight constraint on the parameter space that is compatible with $B$-meson anomalies.  
Similarly as in Fig.~\ref{Combine1}, the $Z'$ masses consistent with $(g-2)_\mu$ are excluded by the bounds from $\tau$ decays and neutrino trident production. 
The LHC bounds constrain the parameter space to the region with a sizable $x$, almost the same for the two cases, as in Fig.~\ref{Combine1}, due to the fact that $BR(Z'\rightarrow\mu^+\mu^-)\sim 30\%$ for $x\ll y$.  The constraints on the mixing parameter of $B_3-L_3$ can be relaxed slightly when one takes into account the acceptance and efficiency of detector~\cite{Alloul:2013bka}. 
We may explore further the remaining parameter space explaining $B$-meson anomalies, by more data from dimuon resonance searches at the LHC.  We can also test our model further from other decay mode such as $\tau{\bar\tau}$ and $\nu_{\mu,\tau}{\bar\nu}_{\mu,\tau}$, the latter of which induces the invisible $Z'$ decay accompanied by jet(s).

\section{Conclusions}
We have proposed a simple $U(1)_{y(L_\mu-L_\tau)+x(B_3-L_3)}$ extension of the SM with favorable couplings to heavy flavors to explain the $B$-meson anomalies. The model requires at least two right-handed neutrinos to cancel the gauge anomalies and gives rise to predictive forms of quark and lepton mass matrices. We have shown that for a small mixing of $B_3-L_3$ in our $U(1)'$, the observed values of $R_K$ and $R_{K^*}$ are consistently accommodated, satisfying bounds from other meson decays/mixings, tau decays, neutrino scattering and electroweak precision data as well as LHC dimuon searches.

\section*{Acknowledgments}

We would like to thank Chengcheng Han, Da Liu, Pyungwon Ko, Yuji Omura, Myeonghun Park, Seong Chan Park and Yong Tang for useful discussions. 
We would also appreciate stimulating discussions with participants during the LHC Physics Workshop on $B$-physics anomalies in Korea Institute for Advanced Study when this work was presented. 
The work is supported in part by Basic Science Research Program through the National Research Foundation of Korea (NRF) funded by the Ministry of Education, Science and Technology (NRF-2016R1A2B4008759). 
The work of LGB
is partially supported by the National Natural Science Foundation of China (under Grant No. 11605016), Korea Research Fellowship Program
through the National Research Foundation of Korea (NRF) funded by the Ministry of Science, ICT and Future Planning (2017H1D3A1A01014046).
The work of SMC is supported in part by TJ Park Science Fellowship of POSCO TJ Park Foundation.
The work of YJK was supported by IBS under the project code, IBS-R018-D1.

\end{document}